\documentclass[aps,english,10pt,a4,superscriptaddress,onecolumn]{revtex4-2}
\AtBeginDocument{\fontsize{11pt}{12.5pt}\selectfont}

\usepackage{amsmath,mathtools}
\usepackage{amsthm}
\usepackage{amssymb}
\usepackage{amsfonts}
\usepackage{bbm}
\usepackage{epsfig}
\usepackage{times}
\usepackage{babel}
\usepackage{color}
\usepackage{hyperref}
\usepackage{mdframed}
\usepackage{changes}
\usepackage{float}
\usepackage{soul}
\usepackage{braket}

\definecolor{shadecolor}{gray}{0.95}

\begin{document}
\title{Quantum Superpositions of Conscious States in a Minimal Integrated Information Model}

\author{Kelvin J.\ McQueen}
\affiliation{Philosophy Department, Chapman University, Orange, CA 92866, USA}
\affiliation{Institute for Quantum Studies, Chapman University, Orange, CA 92866, USA}
\author{Ian T.\ Durham}
\affiliation{Department of Physics, Saint Anselm College, Manchester, NH 03102, USA}
\author{Markus P.\ M\"uller}
\affiliation{Institute for Quantum Optics and Quantum Information, Austrian Academy of Sciences, Boltzmanngasse 3, 1090 Vienna, Austria}
\affiliation{Institute for Quantum Optics and Quantum Information, Austrian Academy of Sciences, 1090 Vienna, Austria}
\affiliation{Perimeter Institute for Theoretical Physics, Waterloo, ON N2L 2Y5, Canada}
\date{March 6, 2026}

\begin{abstract}
Could there be quantum superpositions of conscious states, as suggested by the Wigner’s friend thought experiment? Mathematical theories of consciousness, notably integrated information theory (IIT), make this question more precise by associating physical systems with both quantitative amounts of consciousness and structural characterizations of conscious states. Motivated by a recent proposal that ties wave-function collapse to integrated information, we construct a simple quantum circuit that would, on that proposal, place a minimal system—a feedback dyad—into a superposition of states that differ in their associated conscious states. This “Schr\"odinger’s dyad” provides a controlled setting for evaluating a central desideratum of consciousness-based collapse models: that collapse rates depend on how different the experiences in the superposition are. We prove a structural constraint on collapse dynamics of a standard (Lindblad) type: if collapse is governed by too few collapse operators, collapse rates cannot in general be made to depend solely on qualitative differences between conscious states. Avoiding this limitation requires introducing many commuting operators, leading to a rapid proliferation of collapse terms even for very simple systems. This proliferation bears directly on claims that IIT-based collapse theories may be especially experimentally tractable, since the required dynamics becomes highly complex. More generally, the difficulty is not specific to IIT: any Wigner-style collapse theory that distinguishes experiences using rich internal organization (such as neural connectivity in addition to neural state) will face a comparable explosion in dynamical complexity.
\end{abstract}

\maketitle

\renewcommand \thesection{\arabic{section}}
\renewcommand{\thesubsection}{\arabic{section}.\arabic{subsection}}
\renewcommand{\thesubsubsection}{\arabic{section}.\arabic{subsection}.\arabic{subsubsection}}

\section{Introduction}\label{intro}

Could there be a quantum superposition of conscious states? This question was raised by Eugene Wigner in the thought experiment now known as Wigner’s Friend. He imagined his friend, in a sealed laboratory, performing a quantum measurement while Wigner himself remains ignorant of the outcome. According to the standard quantum description, the laboratory would then be represented by a superposition of different measurement results, which would appear to imply that his friend was in a superposition of experiencing those results. Wigner argued that this was “absurd because it implies that my friend was in a state of suspended animation” and concluded that “consciousness must have a different role in quantum mechanics than the inanimate measuring device” (p.~180, \cite{Wigner1961}). This led him to suggest that consciousness itself might be responsible for wave-function collapse.

Since Wigner’s proposal, there has been extensive debate over whether conscious states could be superposed and how such scenarios should be interpreted. Extensions of the Wigner’s Friend experiment have sharpened the epistemological and consistency issues involved \cite{frauchiger2018quantum, brukner2018no, Durham2019, bong2020strong, wiseman2022thoughtful}, and many-worlds and many-minds interpretations have offered accounts of how superpositions of experiences might be understood \cite{Everett1957, squires1991one, Albert1992, lockwood1996many, Chalmers1996, vaidman1998schizophrenic, barrett1999quantum, lewis2000like, lewis2004many}. However, much of this discussion proceeds without a precise physical criterion for which states are conscious and how different conscious states are to be compared. Without such a criterion, the question of whether conscious states can be superposed and what dynamical consequences this would have remains underdetermined.

Recent neuroscience has seen the development of quantitative theories that aim to relate physical systems to conscious experience. Among these, integrated information theory (IIT) provides an especially explicit mathematical framework, assigning to a system both a quantitative amount of consciousness, denoted $\Phi$, and a structural description of its conscious state (its “$\phi$-structure”) \cite{Tononi2004, Tononi2008, Oizumietal2014, Tononietal2016, albantakis2022integrated}. The neuroscience of consciousness remains a highly contested field with several competing approaches; for a broad overview see \cite{seth2022theories}. IIT has been argued to be consistent with a number of experimental findings \cite{Massiminietal2005, Casarottoetal2016, Haunetal2017, Afrasiabietal2020, Leungetal2020, leung_tsuchiya_2023, ITTAdversarial}. At the same time, substantial criticisms of IIT have been raised \cite{hopkins2022filled, hanson2021formalizing, BarrettMediano2019, Doerig2019, Pautz2019, Bayne2018, Aaronson2014}. Our use of IIT does not depend on its ultimate correctness. Rather, its importance here is methodological: it provides a mathematically precise mapping between physical states and conscious states, making it possible, in principle, to formulate and analyze physical models whose dynamics depend on consciousness.

Once a mathematically explicit characterization of conscious states is available, Wigner’s suggestion can be formulated as a concrete physical proposal. In particular, models have been developed in which wave-function collapse is governed by measures of integrated information associated with a system \cite{KremnizerRanchin2015, ChalmersandMcQueen2022}. The model of \cite{ChalmersandMcQueen2022}, which we analyze here, specifies collapse rates as a function of differences between the conscious states associated with the superposed physical states. Because scalable quantum computations would involve superpositions of distinct conscious states in this sense, the model predicts a breakdown of universal quantum computing and thus bears on recent no-go results in quantum foundations, especially the Local Friendliness theorem \cite{wiseman2022thoughtful}.

A key motivation for such proposals is experimental testability. In comparison to standard collapse models \cite{Bassietal2013}, it has been argued that IIT-based consciousness-collapse models may be experimentally testable with comparatively small systems. In particular, Chalmers and McQueen suggest that suitably designed quantum computers could enter superpositions of conscious states and thereby provide an accessible probe of collapse dynamics \cite{ChalmersandMcQueen2022}. Central to this proposal is a specific dynamical requirement: collapse rates should depend on how different the associated conscious states are. Superpositions corresponding to very different experiences should collapse rapidly, whereas “small” superpositions, involving only minor experiential differences, should persist longer. This requirement is not merely aesthetic. Collapse theories cannot simply impose rapid collapse universally: empirical considerations already pressure such models to allow persistent small superpositions, as reflected in issues such as the tails problem \cite{McQueen2015} and tensions with the quantum Zeno effect \cite{ChalmersMcQueen2024Zeno}. Making collapse rates track differences between conscious states is intended to provide a principled way to satisfy these constraints, and it is this requirement that we analyze in the present paper.

We propose a simple quantum circuit which, if implemented, would place a minimal physical system into a superposition of conscious states, according to the version of IIT used in \cite{ChalmersandMcQueen2022}. This controlled setting allows us to analyze, with a very simple case, what structure a consciousness-based collapse model must incorporate if collapse rates are to track differences between conscious states. In particular, it enables us to assess whether the dynamical complexity required to implement this dependence is compatible with the claim that such models are especially simple or experimentally tractable \cite{eren2026consciousness}.

To make matters concrete, we follow \cite{ChalmersandMcQueen2022} and consider a feedback dyad. Classically, the dyad has four possible states: $(0,0)$, $(1,1)$, $(0,1)$, and $(1,0)$. Each state is predicted by IIT to possess a small but non-zero amount of consciousness, and this feature is stable across successive IIT formalisms. In IIT2.0 each state has $\Phi=2$, in IIT3.0 each has $\Phi=1$~\cite{mcqueen2023parts}, and in Appendix~\ref{4.0Phi} we show that in IIT4.0 each has $\Phi=4$. Although the quantitative value varies, the prediction that the dyad realizes non-zero consciousness is robust, making it a natural minimal test case.

Although these states have the same \emph{amount} of consciousness, the version of IIT employed in \cite{ChalmersandMcQueen2022} assigns them different \emph{states} of consciousness, as reviewed in Section~\ref{Q}. A quantum superposition of two dyad states therefore constitutes, within that model, a superposition of conscious states. We refer to this system as \emph{Schr\"odinger’s dyad} and present a simple quantum circuit that serves as a blueprint for its realization. In Section~\ref{implementation} we describe a possible physical implementation of this circuit, in which two photons enter a feedback loop inside an optical cable.

In Section~\ref{collapse} we analyze collapse dynamics satisfying the requirement that collapse rates depend on differences between the conscious states associated with the superposed basis states. Using the dyad’s conscious-state structure defined in Section~\ref{Q}, we construct a corresponding Lindblad-type collapse model and study its dynamics. Our main result is a structural constraint on such models. If collapse is implemented with too few collapse operators—in particular, with a single operator—then collapse rates cannot, in general, be determined solely by the pairwise qualitative differences between conscious states. The obstruction follows from basic features of quantum dynamics: as we will show, the decay rates depend on the Euclidean distances between tuples of eigenvalues of the $N$ collapse operators, and these distances can reflect the qualitative differences of conscious experience only if $N$ is large enough.

We then show how this limitation can be evaded, and at what cost. Introducing sufficiently many commuting collapse operators can, in principle, reproduce the desired dependence of collapse rates on qualitative differences between conscious states. However, doing so leads to a rapid proliferation of operators even for very simple systems. This bears directly on claims that IIT-based consciousness-collapse models are simple or experimentally tractable, since the complexity of the collapse dynamics is itself part of what determines what an experimental test would involve.

Finally, we show that this complexity does not depend on the particular version of IIT employed in \cite{ChalmersandMcQueen2022}, which uses the IIT3.0 formalism \cite{Oizumietal2014} and the representation of conscious states by “Q-shapes”. The recently proposed IIT4.0 framework \cite{albantakis2022integrated} instead characterizes conscious states in terms of $\phi$-structures. In this formalism the four classical states of the dyad correspond to the same $\phi$-structure and hence to a single conscious state. Nevertheless, as shown in Section~\ref{Phi}, collapse models based on IIT4.0 encounter an analogous difficulty. The reason is that $\phi$-structures include \emph{relations} linking different distinctions in experience, and the number of such relations grows rapidly with system size. For example, the three-node system in Figure~8A of \cite{albantakis2022integrated} contains 60 relations, while the five-node system in Figure~7A contains more than 13,000. A collapse model whose rates track conscious structure must therefore incorporate a corresponding number of dynamical terms, again leading to operator proliferation. We further show that proposed quantum extensions of IIT \cite{zanardi2018quantum, KleinerTull2020, albantakis2023computing} do not remove this constraint and in fact substantially increase the required complexity.

The operator proliferation arises in IIT-based collapse models regardless of which IIT formalism is used. As we will show, this is because conscious states in IIT are individuated by counterfactual causal structure encoded in the system’s transition probability matrix (TPM). The significance of this result may extend beyond IIT. Any theory that individuates experiences by more than the instantaneous state of a physical substrate—for example, by the organization or connectivity of the underlying system in addition to its momentary activation—will introduce multiple independent features that distinguish one experience from another. A collapse dynamics that tracks such experiential differences must therefore contain correspondingly many independent collapse operators. The proliferation we identify may thus reflect a general constraint on consciousness-based collapse proposals.

\section{The Feedback Dyad}\label{ClassicalDyad}

\begin{figure}[H]
\begin{center}
   \begin{tikzpicture}
        %Gate
        \fill[lightgray] (0,0) -- (1.25,0) -- (1.25,1.25) -- (0,1.25) -- (0,0);
        \draw[thick] (0,0) -- (1.25,0) -- (1.25,1.25) -- (0,1.25) -- (0,0);
        %Gate labal
        \node at (0.625,0.625) {SWAP};
        %Inputs
        \draw (-0.5,0.35) -- (0,0.35);
        \draw (-0.5,0.9) -- (0,0.9);
        %Outputs
        \draw (1.25,0.35) -- (1.75,0.35);
        \draw (1.25,0.9) -- (1.75,0.9);
        %Input labels
        \node at (-1,0.35) {B$\;:\;b$};
        \node at (-1,0.9) {A$\;:\;a$};
        %Output labels
        \node at (2.25,0.35) {B$\;:\;a$};
        \node at (2.25,0.9) {A$\;:\;b$};
    \end{tikzpicture}
    \caption{\label{fig:swap} The logical SWAP gate simply exchanges the values $a$ and $b$ of channels A and B respectively such that if the input is (A = a, B = b), then the output is (A = b, B = a).}
\end{center}
\end{figure}
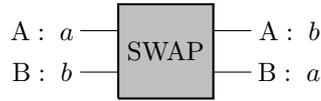

The classical dyad is a simple system consisting of two elements or channels, A and B, that simply swap their states from one time step to the next. That is, if at some time, $t_0$, A is in state 1 and B is in state 0, then at the next time step, $t_{+1}$, A is in state 0 and B is in state 1. The action on these channels is equivalent to a logical SWAP gate, which is given a simple diagrammatic representation in Figure~\ref{fig:swap}.
The figure makes it clear that there are three distinct levels of description to the dyad: channels, channel values, and channel relationships. A and B are the channels that are related via the logical SWAP gate in such a way as to exchange their values. In the language of quantum information, the channels are systems, the channel values are states, and the channel relationships are transformations. The SWAP gate is a transformation of the states of systems A and B. The gate itself is never ``in a state'' on its own.

This distinction matters because in some presentations of IIT---especially in IIT3.0---elements are described in a way that blurs the line between being in a state and implementing a transformation (for example, the abstract of \cite{Oizumietal2014} states that the axioms of IIT are ``formalized into postulates that prescribe how physical mechanisms, such as neurons or logic gates, must be configured to generate experience,” and the first postulate asserts that ``mechanisms in a state exist,” which can be read as treating logic gates themselves as being in states). In the IIT3.0 formalism, nodes are assigned binary values while also being treated as mechanisms that map the states of other nodes to future states. When the elements are understood as neurons, this dual description is often unproblematic: an ``active” or ``inactive” neuron can be naturally interpreted as one that is, or is not, transmitting a signal along a channel \cite{bartlett2022does}. At this level of abstraction, it is convenient to treat neurons as both state-bearers and causal mechanisms. This representational practice has clear precedents in the Boolean network tradition, where nodes are routinely assigned states while also implementing logical update rules \cite{Kauffman1969,Kauffman1990}. However, once one moves to a quantum setting, this ambiguity can no longer be ignored. In quantum theory, superposition applies to the states of physical systems, not to transformations or update rules themselves. Gates implement dynamical relations between systems, but are not entities that can meaningfully be said to be in superposition. For the purposes of the present analysis, we therefore treat the channels (or systems) and their states as quantum, while the SWAP operation is taken to be a fixed transformation acting on those states.

Figure~\ref{fig:swap} also highlights a fundamental causal dependence in the dyad. The output of channel A causally depends on the input to channel B and vice-versa. In order to emphasize this point, we use capital letters to identify the channels or systems themselves and lowercase letters to identify the values the channels can attain, i.e., their states. One could think of the SWAP gate as a black box with the channels simply identifying the locations of the inputs and outputs of the box. Values are fed into the inputs and then produced by the outputs.

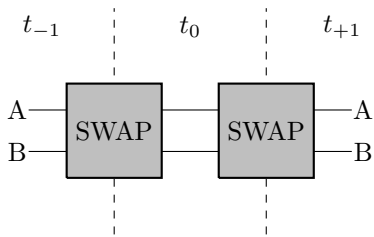
\begin{figure}[H]
\begin{center}
            \begin{tikzpicture}
                %Inputs
                \draw (-0.5,0.35) -- (0,0.35);
                \draw (-0.5,0.9) -- (0,0.9);
                %Input labels
                \node at (-0.65,0.35) {B};
                \node at (-0.65,0.9) {A};
                %Time step t-1
                \node at (-0.325,2) {$t_{-1}$};
                %Time division 1
                \draw[dashed] (0.625,-0.75) -- (0.625,2.25);
                %Gate
                \fill[lightgray] (0,0) -- (1.25,0) -- (1.25,1.25) -- (0,1.25) -- (0,0);
                \draw[thick] (0,0) -- (1.25,0) -- (1.25,1.25) -- (0,1.25) -- (0,0);
                %Gate labal
                \node at (0.625,0.625) {SWAP};
                %Time step t
                \node at (1.625,2) {$t_0$};
                %Outputs-Inputs
                \draw (1.25,0.35) -- (2,0.35);
                \draw (1.25,0.9) -- (2,0.9);
                %Time division 2
                \draw[dashed] (2.625,-0.75) -- (2.625,2.25);
                %Gate
                \fill[lightgray] (2,0) -- (3.25,0) -- (3.25,1.25) -- (2,1.25) -- (2,0);
                \draw[thick] (2,0) -- (3.25,0) -- (3.25,1.25) -- (2,1.25) -- (2,0);
                %Gate label
                \node at (2.625,0.625) {SWAP};
                %Outputs
                \draw (3.25,0.35) -- (3.75,0.35);
                \draw (3.25,0.9) -- (3.75,0.9);
                %Time step t+1
                \node at (3.625,2) {$t_{+1}$};
                %Output labels
                \node at (3.9,0.35) {B};
                \node at (3.9,0.9) {A};
        \end{tikzpicture}
    \caption{\label{fig:dyad} The SWAP gate considered as a feedback system over a series of time steps $t_{-1}$, $t_0$, and $t_{+1}$. The output at any given time step is determined by the input at the previous time step according to the mapping shown in Figure~\ref{fig:swap}.}
    \end{center}
\end{figure}

To develop a feedback system with this SWAP gate we simply feed the outputs directly back into the inputs. For simplicity we can represent this system over a series of time steps in the manner shown in Figure~\ref{fig:dyad}.

The output at a given time step is given as in Figure~\ref{fig:swap}. For example, if the system at a given time step is given by $(a,b)$ with $a,b$ $\in\{0,1\}$ where the first element of the set is the state of channel A and the second is the state of channel B, if the inputs were \mbox{$(a=0,b=1)\equiv(0,1)$,} the evolution of the system state over time is just $(0,1)\to(1,0)\to(0,1)$.

Creating Schr\"{o}dinger's dyad then requires that we treat the channels as quantum and represent their states as such. That is, a classical state $(a,b)$ is equivalent to a pure quantum state in the so-called
computational basis $\ket{a,b}$. A superposition of the $\ket{1,0}$ and $\ket{0,0}$ states can be achieved by feeding the superposition state
\begin{equation}
\ket{+} = \frac{1}{\sqrt{2}}\left(\ket{0}+\ket{1}\right)
\label{plus}
\end{equation}
into channel $B$ at $t_{-1}$. For example, if the input state to the dyad as a whole at $t_{-1}$ is
\begin{equation}
\ket{0,+} = \ket{0} \otimes \ket{+} = \frac{1}{\sqrt{2}}\left(\ket{0,0}+\ket{0,1}\right),
\label{input}
\end{equation} 
then it evolves into the following state at $t_0$:
\begin{equation}
\ket{+,0} = \ket{+} \otimes \ket{0} = \frac{1}{\sqrt{2}}\left(\ket{0,0}+\ket{1,0}\right).
\label{t0state}
\end{equation}

This is not a superposition of $\Phi$ values, since all four possible classical states of the dyad have the same $\Phi$ value. It \textit{is}, however, a superposition of distinct states of consciousness according to the version of IIT used in \cite{ChalmersandMcQueen2022}, as we show in Section~\ref{Q}. The state at time $t_0$ of Equation~(\ref{t0state}) is therefore described as a superposition of qualitatively distinct states of consciousness. To understand this distinction between levels and states of consciousness intuitively, compare yourself experiencing a green screen with experiencing a blue screen. It might be that these two experiences do not correspond to any difference in $\Phi$ (why would changing only the color change the amount of consciousness?). Now imagine that we put a subject into a superposition of experiencing a blue screen and experiencing a green screen. By assumption this is not a $\Phi$ superposition, but it is clearly a superposition of distinct conscious experiences. One might doubt that distinct human states of consciousness would in practice have exactly identical $\Phi$ values \cite{kent2021collapse}, but IIT allows for this in AI (especially in neuromorphic computing), and IIT3.0 and IIT4.0 predict that this is indeed the case for many very simple systems.   

Note that no classical IIT formalism (whether IIT3.0 or IIT4.0) by itself entails this result, since these formalisms describe conscious structure for classical systems and do not specify how such systems behave when placed in quantum superposition. However, classical IIT \textit{does} aspire to say something about the consciousness of \textit{actually existing systems}, in particular those that are well described by classical probability theory, such as neural networks or classical circuits. IIT then inevitably claims non-zero $\Phi$ for some very small or simple actually existing systems. Some of these systems are small and simple enough to be put in superposition states. This will result in superpositions of states that IIT claims are (differently) conscious.

The reasoning here is similar to Schr\"{o}dinger's in the Schr\"{o}dinger's cat thought experiment and to Wigner's in the Wigner's friend thought experiment. But we make no claim about how the superposition of states should be \textit{interpreted}. As an analogy, if we know classically what it means for a spin-1/2 particle to have spin-up or spin-down in the $z$-direction, we can talk about the ``superposition of up and down''. Using these words does not make any claims about how to interpret the resulting quantum state. We need more input from quantum theory (say, group representation theory) to determine that the resulting state actually describes spin pointing up in the $x$-direction. Similarly, perhaps a more general IIT, once properly developed, might say more about the superposition in our case, but that is a question for another day. We now turn to the formal description of the conscious states used in the collapse model in \cite{ChalmersandMcQueen2022}, using the dyad as a simple illustration.

\section{Calculating the States of Consciousness (Q-Shapes) of the Feedback Dyad}\label{Q}

In IIT, the amount of consciousness of a system is quantified by the scalar value $\Phi$, but the theory also associates each physical state with a structural description of its conscious content. In the collapse model of \cite{ChalmersandMcQueen2022}, this qualitative structure is represented by a state's ``Q-shape'', a notion inspired by the IIT3.0 formalism (IIT's mathematical description of conscious states has appeared under several labels. In IIT3.0 \cite{Oizumietal2014} it is primarily called a ``maximally irreducible conceptual structure'' (MICS), and is sometimes described as a ``shape in qualia space'', motivating the simpler term ``Q-shape'' in \cite{ChalmersandMcQueen2022}. In IIT4.0 \cite{albantakis2022integrated} a related role is played by a ``$\phi$-structure'', which we discuss in Section~\ref{Phi}. We note that IIT3.0's formalism admits different interpretive stances regarding whether such structures are defined only up to relabeling symmetries (an intrinsic reading) or whether their explicit coordinate representations can be taken as individuating qualitative states. The collapse model of \cite{ChalmersandMcQueen2022} adopts the latter reading, and our use of ``Q-shape'' follows that proposal. Our main results do not depend on this reading, as shown in Section~\ref{Phi}. The role of Q-shapes in the present paper is methodological: they provide an explicit way of individuating conscious states and comparing them. In this section we construct the Q-shapes for all four classical states of the dyad and show that they are distinct. Within the collapse model, these differences correspond to different conscious states.

The dyad states $(1,0)$ and $(0,0)$ are assigned the same value $\Phi=2$ in IIT3.0 (see \cite{ChalmersandMcQueen2022, mcqueen2023parts} for explicit calculations), yet in the collapse model in \cite{ChalmersandMcQueen2022} they are treated as qualitatively different. The difference arises when the system is partitioned—i.e., when certain causal connections are replaced by noise—and one examines the probability distributions obtained by evolving the partitioned system forward and backward in time. These distributions are not the same for $(1,0)$ and $(0,0)$, and the Q-shape records this difference. Thus the Q-shape distinguishes dyad states even when their $\Phi$ values coincide. In more complex systems, the contributions of different subsystems must also be weighted by their integrated information $\phi$. In the dyad, however, $\phi(A)=\phi(B)=1$, so no additional weighting is required and the qualitative structure can be represented directly by simple matrices.

We begin with part A, when the dyad is in state $(1,0)$. The prescription of \textit{partitioning} means that we replace the complement of A (that is, B) by noise, i.e.,\ an equiprobable distribution of $0$ and $1$, while keeping $A$ in state $1$. Evolving this forward in time, we obtain a probability distribution $(0,\frac 1 2,0,\frac 1 2)$, where we have labelled the four states in lexicographical order: $(0,0),(0,1),(1,0),(1,1)$. Computing the corresponding backward (cause) distribution over the previous time step yields exactly the same probability distribution. This gives us the first two rows in the Q-shape matrix
\begin{equation}
   Q(1,0)=\left(\begin{array}{cccc}
	0 & \frac 1 2 & 0 & \frac 1 2 \\
	0 & \frac 1 2 & 0 & \frac 1 2 \\
	\frac 1 2 & \frac 1 2  & 0 & 0 \\
	\frac 1 2 & \frac 1 2  & 0 & 0
\end{array}
   \right).
   \label{eqQShape1}
\end{equation}
The third and fourth rows are the forward (effect) and backward (cause) probability distributions that we obtain if we consider the subsystem B instead, keeping it in state $0$ and replacing A by noise as above. Thus, the Q-shape of a given state (such as $(1,0)$) is a collection of four probability distributions over the four dyad states, represented by the four rows in our representation matrix.

Performing the calculation for the other dyad states (which have identical $\Phi$), \mbox{we obtain}
\begin{equation} 
   Q(0,0)=\left(\begin{array}{cccc}
	\frac 1 2  & 0 & \frac 1 2 & 0 \\
	\frac 1 2  & 0 & \frac 1 2 & 0 \\
	\frac 1 2 & \frac 1 2  & 0 & 0 \\
	\frac 1 2 & \frac 1 2  & 0 & 0
\end{array}
   \right),\enspace
      Q(0,1)=\left(\begin{array}{cccc}
	\frac 1 2  & 0 & \frac 1 2 & 0 \\
	\frac 1 2  & 0 & \frac 1 2 & 0 \\
	0 & 0  & \frac 1 2 & \frac 1 2 \\
	0 & 0  & \frac 1 2 & \frac 1 2
\end{array}
   \right),\enspace
   Q(1,1)=\left(\begin{array}{cccc}
	0 & \frac 1 2 & 0 & \frac 1 2 \\
	0 & \frac 1 2 & 0 & \frac 1 2 \\
	0 & 0  & \frac 1 2 & \frac 1 2 \\
	0 & 0  & \frac 1 2 & \frac 1 2
\end{array}
   \right).
   \label{eqQShape2}
\end{equation}

Our Q-shapes are not literal geometric “shapes”; they are matrices encoding the forward and backward probability distributions associated with each subsystem. The terminology is motivated by the IIT3.0 visualizations in \cite{Oizumietal2014} (see especially \mbox{Figures~10--12}), where qualitative structure is depicted as a configuration in a high-dimensional space. Rather than working with the geometric picture, we follow \cite{ChalmersandMcQueen2022} and use an explicit coordinate representation of the same information.

In the dyad, each mechanism (the single-element subsystems $A$ and $B$) contributes a \emph{pair} of probability distributions over the four system states: its effect distribution and its cause distribution. Concatenating these two four-component distributions yields a vector in $\mathbb{R}^8$. For $Q(1,0)$ in Equation~(\ref{eqQShape1}), the first two rows (effect and cause for $A$) determine the point associated with $A$, and the last two rows determine the point associated with $B$. The Q-shape of a dyad state can therefore be represented by the pair of such points.

In full IIT3.0, these structures (or ``concepts'') are additionally weighted by their integrated-information values $\phi$. In the dyad, however, $\phi(A)=\phi(B)=1$ for all four classical states, so the weights do not differentiate the Q-shapes. Accordingly, we suppress these uniform weights and represent each dyad Q-shape solely by the associated \mbox{probability distributions.}

IIT predicts that the dyad has a non-zero value of $\Phi$ and hence, within the theory, qualifies as minimally conscious. In the collapse model of \cite{ChalmersandMcQueen2022}, the four dyad states are further treated as corresponding to distinct qualitative states via their Q-shapes. It is therefore natural to ask what, if anything, such a system might experience \cite{nagel1980like,Tsuchiya2017}. We do not pursue that question here. For comparison, (Figure~19, \cite{Oizumietal2014}) discusses the possible experience of a simple photodiode system with non-zero $\Phi$. Whether or not such interpretations are ultimately correct is not important for our purposes. The present analysis concerns only the mathematical representation of conscious states and the dynamical constraints that follow if collapse rates depend on differences between them.

Before turning to collapse dynamics, we briefly discuss how a system of the kind just described might be physically instantiated. The next section therefore considers a simple optical implementation of the dyad and how the relevant causal units should be identified.

\section{Physically Implementing the Dyad}\label{implementation}

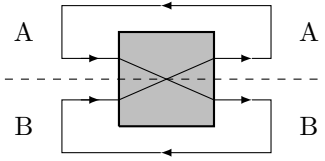
\begin{figure}[H]
\begin{center}
    \begin{tikzpicture}
        %Gate
        \fill[lightgray] (0,0) -- (1.25,0) -- (1.25,1.25) -- (0,1.25) -- (0,0);
        \draw[thick] (0,0) -- (1.25,0) -- (1.25,1.25) -- (0,1.25) -- (0,0);
        %Inputs
        \draw (-0.5,0.35) -- (0,0.35);
        \draw (-0.5,0.9) -- (0,0.9);
        %Outputs
        \draw[-latex] (1.25,0.35) -- (1.75,0.35);
        \draw[-latex] (1.25,0.9) -- (1.75,0.9);
        %Internal crossing
        \draw (0,0.35) -- (1.25,0.9);
        \draw (0,0.9) -- (1.25,0.35);
        %Feedback loops
        \draw[-latex] (1.75,0.35) -- (2,0.35) -- (2,-0.35) -- (0.55,-0.35);
        \draw[-latex] (0.55,-0.35) -- (-0.75,-0.35) -- (-0.75,0.35) -- (-0.25,0.35);
        \draw[-latex] (1.75,0.9) -- (2,0.9) -- (2,1.6) -- (0.55,1.6);
        \draw[-latex] (0.55,1.6) -- (-0.75,1.6) -- (-0.75,0.9) -- (-0.25,0.9);
        %Qubits
        \draw[dashed] (-1.5,0.625) -- (2.75,0.625);
        \node at (-1.25,1.25) {A};
        \node at (-1.25,0) {B};
        \node at (2.5,1.25) {A};
        \node at (2.5,0) {B};
    \end{tikzpicture}
\caption{%
\label{fig:implementation}
A possible
 implementation of the dyad. The dashed line and labels indicate that the systems A and B are associated with regions of space.
}
\end{center}
\end{figure}

Consider the following simple implementation of the dyad depicted in Figure \ref{fig:implementation}: channels A and B are optical cables and the dyad does nothing more than cross those cables, without contact. The outputs are then fed back into the inputs, creating a kind of feedback cycle. We have two photons in the cables, and each of them can carry one of two perfectly distinguishable states, corresponding to horizontal $|0\rangle$ or vertical polarization $|1\rangle$. What horizontal or vertical means is determined by an external reference frame; for what follows, the exact choice of reference is unimportant, except that the physical situation must tell us what we mean by both photons carrying \emph{identical} or \emph{orthogonal} polarization directions (e.g.,\ $|00\rangle$ in the first case, and $|01\rangle$ in the second). It is clear that we are not restricted to preparing the photons in these basis states, but we can prepare them in arbitrary superpositions, such as that of Equation~(\ref{t0state}). This is a necessary condition to implement ``Schr\"odinger's dyad'' as introduced in the previous sections.

However, a subtlety arises concerning what should count as the basic causal units of the system. One possibility is to identify the units with the photons themselves as they traverse the cables. On this interpretation, each photon’s future polarization depends only on its own prior polarization and not on the state of the other photon. The joint system therefore factorizes into two independent subsystems. In IIT terms, each photon may possess intrinsic cause–effect power, but the pair does not form an integrated system: the composite system is reducible to the separate evolution of its parts. Under this description the setup would not implement a dyad, but rather two independent identity channels.

On the other hand, this depends on counting the photons as our basic causal units. If we instead identify our basic units as the polarization qubits at the physical locations $A$ and $B$ in space, we get a different result. In particular, we may say that the photon polarization state at A$_{t_0}$ causes the state at B$_{t_{+1}}$ and was caused by the state at B$_{t_{-1}}$. Under this way of identifying our basic units, the system has non-zero $\Phi$ and implements a very simple conscious system according to IIT. 

In IIT, the $\Phi$ of a system is meant to be an objective property, so should not depend on some arbitrary choice. So at least one of the above two causal interpretations of the system must be ruled out. We think that the IIT4.0 \textit{principle of maximal existence}, which states that ``what exists is what exists the most'', can help here. As stated in \cite{albantakis2022integrated}, ``By the maximal existence principle, the complex should be the one that lays the greatest claim to existence as one entity, as measured by system integrated information.'' The principle is a criterion for choosing the right complex of units (among several complexes that overlap on a given substrate), which is the relevant question for us. The principle is used to motivate the IIT exclusion principle, which effectively states that if two overlapping sets of units have non-zero $\Phi$, then only the system with \textit{maximal} $\Phi$ is conscious. But if there are multiple interpretations of what the causal units are in the first place, we might similarly only consider the interpretation that yields the greater $\Phi$. Applied to the present case, this principle favors treating the polarization degrees of freedom at spatial locations A and B as the relevant causal units, since this identification yields a system with non-zero $\Phi$ and a nontrivial causal structure. Adopting this perspective suffices for our purposes here: it provides a physically well-motivated implementation of the dyad to which the IIT-based collapse analysis of Section~\ref{collapse} can be applied.

\section{A Simple Consciousness-Collapse Model}\label{collapse}

In \cite{ChalmersandMcQueen2022}, a dynamical collapse model is proposed in which superpositions of distinct Q-shapes (qualitative conscious states) are dynamically unstable and tend to collapse. The goal was to formulate a consciousness-collapse model that is at least as rigorous as standard collapse models \cite{Bassietal2013}. The authors conclude that their resulting model ``is not as simple or powerful as some of the leading interpretations [\ldots]
 Nevertheless, it at least serves as an existence proof for a relatively precise consciousness-collapse model'' (p.~50, \cite{ChalmersandMcQueen2022}). The model has since been discussed as a potential response to the Local Friendliness theorem \cite{wiseman2022thoughtful}, which forces the introduction of a new assumption that the model denies, an assumption Wiseman et al.\ call ``Universal Quantum Computing'' \cite{wiseman2022thoughtful}. The model also shows how the IIT formalism can be extended to different types of physical theories, in this case to a theory with collapse.

The claim that IIT-based consciousness-collapse models are comparatively \emph{simple} or \emph{experimentally tractable} deserves careful scrutiny. In this section we identify a structural constraint on a broad class of Lindblad-form collapse models: if collapse is governed by too few collapse operators---most notably, a single operator---then the pairwise collapse rates between distinct pairs of conscious states cannot, in general, be made to depend \emph{only} on their pairwise qualitative differences. Conversely, if one introduces enough collapse operators to recover the desired dependence on qualitative differences, one obtains a rapid proliferation of operators even for very simple systems such as the dyad. This yields \mbox{a dilemma:}
\begin{quote}
\emph{Either} the model uses few collapse operators, in which case collapse rates fail to track conscious structure in the intended pairwise way; \emph{or} it uses many collapse operators, in which case the dynamics becomes surprisingly complex even for small systems.
\end{quote}

To establish the first horn of the dilemma, we define a simple collapse model which has just one collapse operator, whose eigenstates correspond to total states of consciousness, i.e.,\ Q-shapes. For such a model, we prove that it is impossible to choose the eigenvalues of a single collapse operator so that, for every pair of basis states, the collapse rate of their superposition is large if and only if the corresponding states of consciousness are qualitatively very different.
While our formal result concerns the case of a single collapse operator, the underlying source of the problem---the mismatch between the number of collapse parameters and the number of pairwise qualitative distances---suggests that similar constraints will arise whenever collapse is governed by too few operators. As we show below, this difficulty can be avoided by introducing additional collapse operators, but at the cost of a rapid increase in dynamical complexity, even for the simple dyad.

The model in \cite{ChalmersandMcQueen2022} uses the following general form for continuous collapse \mbox{models (p.27, \cite{Bassietal2017}):}

\begin{equation}\label{Bassi(74)}
d\psi_t = [-i\hat{H}_0dt + \sqrt{\omega}(\hat{A} - \langle\hat{A}\rangle_t)dW_t - \frac{\omega}{2}(\hat{A}-\langle\hat{A}\rangle_t)^2dt ]\psi_t.
\end{equation}
The first term on the right-hand side of the equation represents Schr\"odinger evolution, while the remaining two terms represent the collapse evolution. Here, $\hat{H}_0$ is the Hamiltonian of the system, $\omega$ is a real-valued parameter governing the collapse rate, $\hat{A}$ is a collapse operator whose eigenstates the system collapses towards, $\langle\hat{A}\rangle_t$ is its expected value at time $t$, and $W_t$ is a noise process which ensures that collapse happens stochastically at a rate determined by a measure of difference between the superposed $\hat{A}$ eigenstates.

The pure state $\rho_t^W:=|\psi_t\rangle\langle\psi_t|$ therefore evolves stochastically. The totality of all predictions that quantum physics makes for measurement outcomes in experiments is given by the expectation values $\langle\psi_t|\hat O|\psi_t\rangle={\rm tr}(\rho_t^W\hat O)$, where $\hat O =\hat O^\dagger$ is any observable and $\rho_t^W:=|\psi_t \rangle\langle\psi_t|$ is the density operator that corresponds to $\psi_t$. However, we have no access to the actual realization of the noise $W_t$ and the resulting pure-state trajectory. Hence, the best possible prediction of measurement outcomes (and, equivalently, the statistics we will record on repeating the corresponding experiment a large number of times) will correspond to the expectation values $\langle\hat O\rangle:=\mathbb{E}(\langle\psi_t|\hat O|\psi_t\rangle)={\rm tr}(\rho_t \hat O)$, where $\rho_t:=\mathbb{E}(\rho_t^W)$ is the density operator averaged over all realizations of the noise process. As a consequence of~(\ref{Bassi(74)}), this resulting state evolves according to the Lindblad equation
\begin{equation}
   \frac d {dt} \rho_t = -i [\hat H_0,\rho_t] -\frac\omega 2 [\hat A,[\hat A,\rho_t]]
   \label{eqLindblad}
\end{equation}
(for the derivation see, e.g.,~\cite{Bassietal2017}). The system can evolve via Schr\"{o}dinger dynamics, via collapse, or via some combination of the two. To understand the collapse term we can ignore the Schr\"{o}dinger dynamics term by setting its Hamiltonian to zero, $\hat H_0=0$. The collapse term only has an effect when the system is in a superposition of eigenstates of $\hat A$. In this situation, the double commutator will be non-zero and the state will evolve. The ``speed'' at which it evolves is a function of the eigenvalues $a_i$ of $\hat A$. This is because the $(i,k)$th matrix entry of the double commutator in $\hat A$'s eigenbasis is

\begin{equation}
   [\hat A,[\hat A,\rho]]_{ik}=\rho_{ik}(a_i-a_k)^2.
\end{equation}

The dampening of the off-diagonal elements of $\rho$ occurs at a rate that grows with $(a_i-a_k)^2$ where if $a_i \ne a_k$ the system is in a superposition. We see that the eigenbasis of $\hat A$ determines the collapse basis, i.e.,\ the basis in which the state becomes ``classical'', while its eigenvalues tell us which superpositions of pairs of such states are removed more quickly (namely, those with large $(a_i-a_k)^2$). Let us now use this prescription to construct a very simple consciousness-collapse model for the dyad, involving only a single collapse operator.

The four states of the dyad correspond to mutually distinct states of consciousness and span the Hilbert space of the system. A consciousness-collapse model should therefore drive the state, at large times $t$, to one that is diagonal in this basis (this is the preferred basis for the purpose of our simple example. Note that all collapse models, including standard collapse models, postulate a preferred basis. For GRW, it is one whose basis states are definite positions states. For CSL, it is one whose basis states are definite mass-density states. For consciousness-collapse models, it is one whose basis states are definite states of consciousness. For all such models, it is one whose basis states the state vector collapses to. For the purpose of modeling the dyad, we assume that the computational basis yields definite states of consciousness. But for consciousness-collapse models generally, the preferred basis need not correspond exactly to a ``classical basis''). Therefore, our collapse operator $\hat A$ will have the form

\begin{equation}
   \hat A=\lambda_{00}|00\rangle\langle 00|+\lambda_{01}|01\rangle\langle 01|+\lambda_{10}|10\rangle\langle 10|+\lambda_{11}|11\rangle\langle 11|,
\end{equation}

with four eigenvalues $\lambda_{ij}$. 
Any consciousness-collapse model should arguably imply the following principle for the choice of those eigenvalues:

\textit{Superpositions of two dyad states (say, $ij$ and $kl$) should vanish very quickly if and only if they are qualitatively very different states of consciousness.}

That is, it is natural to allow superpositions of ``qualitatively similar'' states to persist for longer, while qualitatively different states must decohere quickly. As mentioned in the introduction, this principle is motivated by familiar constraints on collapse dynamics—such as the need to avoid tails problems and pathological Zeno-type behavior—which push against indiscriminate or uniformly rapid collapse

For a quantitative application of this prescription, we need a way to compare states of consciousness, i.e.,\ a distance measure on Q-shapes (Figure 11, \cite{Oizumietal2014}). Let us first begin with a simple ansatz that will help us to get some intuition for what is going on, before coming to a more general and rigorous no-go result further below.

\textbf{Intuitive ansatz.} Since Q-shapes are collections of probability distributions, it is natural to define their distance in terms of distance measures on probability distributions, which is a classical and well-studied topic in information theory. The preferred distance measures on probability distributions in IIT have changed in almost every successive version. IIT2.0 used the well-known Kullback--Leibler divergence. IIT3.0 used Earth Mover's distance \cite{pele2009fast}. IIT4.0 uses the intrinsic difference measure described in Appendix \ref{4.0Phi}.

A natural choice is to define the distance of two Q-shapes $Q=(q_1,q_2,q_3,q_4)^\top$ (i.e.,\ with rows $q_1,\ldots,q_4$) and $\tilde Q=(\tilde q_1,\tilde q_2,\tilde q_3,\tilde q_4)^\top$ as
\begin{equation}
   \mathcal{D}(Q,\tilde Q):=\sum_{i=1}^4 \mathcal{D}(q_i,\tilde q_i),
   \label{Qdistance}
\end{equation}
where $\mathcal{D}$ is some choice of distance measure on the set of probability distributions. That is, the distance of two Q-shapes is the sum of the distances of their probability distributions. (This is very similar to the form of IIT3.0's extended Earth mover's distance measure.)

Now we have a large choice of possible distance measures $\mathcal{D}$ at our disposal. However, note that the four Q-shapes of the dyad (Equations~(\ref{eqQShape1}) and~(\ref{eqQShape2})) consist of a small variety of very simple probability distributions only: all entries are $0$ or $\frac 1 2$, and any two rows are either equal, or they differ in all four entries. Two identical rows must have distance zero. Furthermore, it is natural to demand that every two probability distributions arising as rows in these Q-shapes that differ in all four places all have the same distance, which we can set to unity by a choice of scaling factor. For example,
\[ \textstyle
   \mathcal{D}\left(\strut (0,\frac 1 2,0,\frac 1 2),(\frac 1 2,0,\frac 1 2,0)\right)=1.
\]

We can then determine the distances between all pairs of Q-shapes of the dyad and obtain the following values, writing $\mathcal{D}(Q,\tilde Q)$ as the $Q\tilde Q$-entry of a table:

\begin{center}
\begin{tabular}{c|cccc|} 
 & Q(0,0) & Q(0,1) & Q(1,0) & Q(1,1)\\
 \hline
 Q(0,0) & $0$ & $2$ & $2$ & $4$ \\
 Q(0,1) & $2$ & $0$ & $4$ & $2$ \\
 Q(1,0) & $2$ & $4$ & $0$ & $2$ \\
 Q(1,1) & $4$ & $2$ & $2$ & $0$
\end{tabular}
\end{center}

Let us now return to our consciousness-collapse principle, and let us focus only on one implication of its if-and-only-if statement. It reads: \textit{If the distance $\mathcal{D}$ between two Q-shapes $Q(i,j)$ and $Q(k,l)$ is large, then the distance between the eigenvalues $\lambda_{ij}$ and $\lambda_{kl}$ of the collapse operator must also be large.}

This desideratum could always be satisfied by the arbitrary prescription to make all eigenvalues extremely large and distant from each other. However, this would typically induce almost-instantaneous collapse, a behavior that we do not expect for simple systems such as the dyad. Thus, we are searching for a choice of eigenvalues that is as tame as possible while still satisfying the above postulate.

There are many different ways to formulate the requirement ``the four eigenvalues $\lambda_{ij}$ are not very large'' as a precise mathematical condition. Since we are still at the stage of developing some intuition (and not yet proving a rigorous no-go result, which will follow below), let us choose one possible formulation arbitrarily which will turn out to lead to a tractable solution: that the sum of all the $\lambda_{ij}$ is small.
This leads us to define the eigenvalues in terms of an optimization problem:\\

\fbox{\begin{minipage}{25em}
Minimize $\lambda_{00}+\lambda_{01}+\lambda_{10}+\lambda_{11}$

subject to $\lambda_{ij}\geq 0$,\quad $|\lambda_{ij}-\lambda_{kl}|\geq \mathcal{D}(Q(i,j),Q(k,l))$.
\end{minipage}}\\

This prescription keeps the collapse behavior ``tame'' by demanding that the eigenvalues are not arbitrarily large, but only as large as they need to be (in their total sum) to satisfy our principle for all pairs of Q-shapes. Note that the total time scale of the collapse is not determined by $\hat A$ and its eigenvalues, which do not have any physical units. Instead, it is determined by the noise term of~(\ref{Bassi(74)}), i.e.,\ the parameter $\omega$ in~(\ref{eqLindblad}). This will remain a parameter of the collapse model that needs to be determined experimentally. The above considerations tell us only the \textit{relative} speed at which superpositions between distinct Q-shapes are suppressed, whereas the \textit{total} speed would depend on $\omega$ and hence on further considerations as to which states of consciousness are implausible to remain in superposition for significant amounts of time because of, say, human experience.

As we show in Appendix~\ref{optimal}, this optimization problem has eight solutions:
\[
  \begin{pmatrix} \lambda_{00} \\ \lambda_{01} \\ \lambda_{10} \\ \lambda_{11}\end{pmatrix}\in\left\{\begin{pmatrix}0 \\ 2 \\ 6 \\ 4\end{pmatrix},\begin{pmatrix} 0\\6\\2\\4\end{pmatrix},\begin{pmatrix}2\\0\\4\\6\end{pmatrix},\begin{pmatrix}6\\0\\4\\2\end{pmatrix},\begin{pmatrix}6\\4\\0\\2\end{pmatrix},\begin{pmatrix}2\\4\\0\\6\end{pmatrix},\begin{pmatrix}4\\2\\6\\0\end{pmatrix},\begin{pmatrix}4\\6\\2\\0\end{pmatrix}\right\}.
\]
This degeneracy can be understood as a consequence of the symmetry of the problem: for example, the table of pairwise distances does not change if we exchange $Q(0,0)$ and $Q(1,1)$, or $Q(0,1)$ and $Q(1,0)$.

We can simply pick one of the six solutions and use it to define our collapse operator. For the sake of argument, let us pick $(\lambda_{00},\lambda_{01},\lambda_{10},\lambda_{11})=(2,0,4,6)$ (though the choice does not matter for the following discussion).

Let us interpret the result by looking at some example collapse rates. We have $\mathcal{D}(Q(0,0),Q(0,1))=2$ which is small, and $|\lambda_{00}-\lambda_{01}|=2$ which is also small (and, indeed, identical). Superpositions of the two dyad states $(0,0)$ and $(0,1)$ can thus remain stable for a relatively long time. On the other hand, $\mathcal{D}(Q(0,1),Q(1,0))=4$ is large, and so is $|\lambda_{01}-\lambda_{10}|=4$. Hence, superpositions between the dyad states $(0,1)$ and $(1,0)$ will be killed off more quickly.

{However, consider the two dyad states {$(0,1)$ and $(1,1)$.} Their distance, \linebreak  \mbox{$\mathcal{D}(Q(0,1),Q(1,1))=2$}, is small and our principle demands that the corresponding difference of eigenvalues (i.e.,\ the associated collapse rate) is at least as large as that. However, it is actually $|\lambda_{01}-\lambda_{11}|=6$, which is much larger than required. Thus, any superposition of these two dyad states would fall off much faster than what would be expected by considering the difference between their Q-shapes alone.}

\textbf{A rigorous no-go result.} One might object that our calculation above is not fully general. For example, we might have a distance measure that is not of the form~(\ref{Qdistance}) (for example the Euclidean distance), and the eigenvalue differences might not be directly related to the distance $\mathcal{D}(Q,\tilde Q)$ between Q-shapes $Q$ and $\tilde Q$, but to some \textit{function} thereof. Furthermore, minimizing the sum of the $\lambda_{ij}$ was a pretty arbitrary choice within the mathematical model. Let us therefore show a more rigorous and general impossibility result. A larger class of natural distance measures $\mathcal{D}$ consists of those that are sensitive (only) to the number of entries on which our $Q(i,j)$ and $Q(k,l)$ differ, giving larger values if more entries are different. In this case, there are two positive numbers $a$ and $b$ with $a<b$ such that
\begin{eqnarray*}
a&=&\mathcal{D}(Q(0,0),Q(0,1))=\mathcal{D}(Q(0,0),Q(1,0))=\mathcal{D}(Q(0,1),Q(1,1)),\\
b&=&\mathcal{D}(Q(0,0),Q(1,1))=\mathcal{D}(Q(0,1),Q(1,0)).
\end{eqnarray*}
This includes the cases discussed above, where $a=2$ and $b=4$, but it also includes the Euclidean distance in $\mathbb{R}^{16}$, where $a=\sqrt{2}$ and $b=2$, or the $p$-norm distance for $p\geq 1$, given by $\mathcal{D}(A,B):=\|A-B\|_p:=\left(\sum_{m,n=1}^4 |A_{mn}-B_{mn}|^p\right)^{1/p}$, yielding $a=\frac 1 2\cdot 8^{1/p}$ and $b=\frac 1 2 \cdot 16^{1/p}$, reducing to the Euclidean distance for $p=2$.

A natural desideratum for our collapse model would be that \textit{there is some distance measure $\mathcal{D}$ such that the collapse rates, determined by $|\lambda_{ij}-\lambda_{kl}|$, are larger iff the corresponding pair of Q-shapes has larger $\mathcal{D}(Q(i,j),Q(k,l))$}.

We will now show that this is impossible for the natural class of distance measures that we allow. To satisfy the requirement, we would have a strictly increasing function $f:\mathbb{R}_0^+\to\mathbb{R}_0^+$ with $f(0)=0$ such that $|\lambda_{ij}-\lambda_{kl}|=f(\mathcal{D}(Q(i,j),Q(k,l))$. Since only the \textit{differences} of the eigenvalues matter for the collapse rates, we can without loss of generality assume that one of the $\lambda_{ij}$ is zero. For example, suppose that $\lambda_{00}=0$. Then, on the \mbox{one hand,}
\vspace{-6pt}
\begin{eqnarray*}
\lambda_{01}&=&|\lambda_{01}-\lambda_{00}|=f(\mathcal{D}(Q(0,1),Q(0,0))=f(a),\\
\lambda_{10}&=&|\lambda_{10}-\lambda_{00}|=f(\mathcal{D}(Q(1,0),Q(0,0))=f(a),
\end{eqnarray*}
and on the other hand, $0=|\lambda_{01}-\lambda_{10}|=f(\mathcal{D}(Q(0,1),Q(1,0))=f(b)>f(a)>0$, which is a contradiction. We obtain similar contradictions in the other three cases of $\lambda_{ij}=0$.

Thus, for our case of a single collapse operator $\hat A$, every value of $|\lambda_{ij}-\lambda_{kl}|$ must depend on more than just the number $\mathcal{D}(Q(i,j),Q(k,l))$. If our principle is satisfied, then a large value of the latter implies a large value of the former, but the converse is not in general true. We must hence have pairs of Q-shapes whose superposition must collapse more quickly than what their mere qualitative distance as states of consciousness would suggest. Superposition resistance hence cannot only reflect the structure of conscious experience, but is also additionally constrained by the general structure of quantum mechanics.

This obstruction can be avoided by having a very large number of collapse operators---essentially, one per independent component of the Q-shapes: Lindblad-form collapse dynamics with multiple commuting collapse operators $\{\hat A_m\}$ suppress off-diagonal terms according to
\vspace{-6pt}
\begin{equation}
\sum_{m=1}^M [\hat A_m,[\hat A_m,\rho]]_{ik}
=
\rho_{ik}\sum_{m=1}^M\bigl(a_i^{(m)}-a_k^{(m)}\bigr)^2=\rho_{ik}\|\vec a_i-\vec a_k\|^2,
\label{multiLindblad}
\end{equation}
replacing the right-hand side of ~(\ref{eqLindblad}), where $\vec a_i=(a_i^{(1)},a_i^{(2)},\ldots,a_i^{(M)})$ is a vector containing the $i$th eigenvalue of all $M$ collapse operators $\hat A_m$. Consider all relevant Q-shapes of a given system, and the affine-linear space that is spanned by them, and let $M$ be the dimension of this space. In a nutshell, $M$ counts the number of (affinely) independent components of the Q-shapes. For example, in our dyad example, the Q-shapes $Q(i,j)$ are elements of $\mathbb{R}^{16}$ (or, equivalently, the space of $4\times 4$ matrices), but they form the four vertices of a tetrahedron, which spans a three-dimensional affine subspace. If $\vec a_k$ denotes the location of the $k$th Q-shape $Q_k$ in this space, then $\|\vec a_i-\vec a_k\|^2$ is the squared Euclidean distance between the Q-shapes $Q_i$ and $Q_k$. Hence, by choosing the $\hat A_m$ as diagonal operators where the $i$th eigenvalue is given by $a_i^{(m)}$, we obtain a collapse model where the collapse rates are exactly identical (up to a global scaling factor) to the distances between the Q-shapes.

This result, however, relies crucially on the fact that one is fitting distances on a finite set of four conscious states. Nothing in this construction singles out the collapse operators in a principled, system-independent way, nor does it generalize to larger systems whose qualitative structure varies along many independent dimensions. As the number of distinguishable conscious states increases, the dimension required to realize the full pairwise distance structure grows rapidly, and no fixed small set of collapse operators can be expected to suffice in general.

To obtain a collapse model that faithfully tracks qualitative differences across all possible conscious states, one must instead associate collapse operators directly with the independent components of a Q-shape. In this case, each collapse operator corresponds to a particular Q-shape component, and the collapse rate between two states is given by the squared Euclidean distance between their Q-shape representations, as made explicit by Equation~(\ref{multiLindblad}). This guarantees that superpositions collapse quickly if and only if the corresponding conscious states are qualitatively very different, something we have shown to be impossible with a single collapse operator.

The cost of this fidelity is substantial dynamical complexity. For a classical IIT system with $n$ elements, each having $d$ possible states, there are $2^n-1$ subsystems. Each subsystem is associated with two probability distributions (cause and effect repertoires) over $d^n$ possible system states, as well as a single $\phi$-value. Capturing the full Q-shape structure may therefore require
\[
(2^n-1)\times(2d^n+1)
\]
independent collapse operators, unless one obtains some accidental affine-linear dependencies. For the classical dyad ($n=d=2$), this yields $(4-1)\times(2\cdot4+1)=27$ collapse operators (strictly speaking, if we employ the simplifications employed in Section \ref{Q}, we could reduce this to $n \times (2 \times d^n) = 16$ collapse operators (one for each matrix entry e.g., in Equation \eqref{eqQShape1}). But these simplifications were possible only because we already knew that for the dyad, the composite subsystem AB has zero $\phi$ (so can be ignored) while the two elements have the same $\phi$ no matter the dyad's state (so can also be ignored). A model for a more general class of systems cannot adopt these simplifications). Even modest increases in system size lead to rapid growth: for $n=d=3$, one already requires $385$ operators.

Importantly, this complexity is not an artifact of restricting attention to classical IIT. In the consciousness-collapse model of \cite{ChalmersandMcQueen2022}, the Q-shapes are defined using quantum integrated information theory \cite{zanardi2018quantum,KleinerTull2020}, where each element is a qubit with a continuum of possible pure states. In this case, each subsystem is associated with two density matrices rather than two classical probability distributions. Since a density matrix on $d^n$ dimensions has $d^{2n}$ independent real parameters, the number of collapse operators required becomes
\[
(2^n-1)\times(2d^{2n}+1).
\]
For the quantum dyad, this already yields $(4-1)\times(2\cdot16+1)=99$ collapse operators.

The source of the complexity identified here ultimately arises from the fact that IIT defines conscious structure in terms of counterfactual causal relations. If collapse rates are required to track differences in Q-shapes, then collapse dynamics must depend not only on the system's physical actual state, but also on how that system would behave under a wide range of counterfactual interventions. Standard collapse models such as CSL do not face this burden: their collapse operators depend only on actual physical observables (such as mass density), and make no reference to counterfactual state transitions. The resulting operator proliferation in IIT-based collapse models is therefore not an artifact of modeling choices, but a direct consequence of the counterfactual content of the underlying theory \mbox{of consciousness.}

The resulting picture is therefore fourfold. With a single collapse operator, collapse rates cannot track qualitative differences in conscious states. With a small, system-specific set of operators, one can sometimes fit the behavior of particular finite systems, such as the dyad, but without principled generality. With one operator per independent classical Q-shape component, collapse faithfully tracks conscious structure at the cost of extreme complexity. Finally, quantum extensions of IIT only exacerbate this burden. The dilemma identified here is thus robust: the apparent complexity of consciousness-collapse models is not an optional modeling choice, but a structural consequence of requiring collapse dynamics to reflect qualitative differences in conscious experience.

\section{Extending the Result to IIT4.0 and Beyond}\label{Phi}

The previous analysis employed the IIT3.0-inspired notion of a Q-shape because it provides a simple and explicit representation of qualitative differences between conscious states. Since IIT has been reformulated as IIT4.0 \cite{albantakis2022integrated}, with conscious states described now by $\phi$-structures, it is important to determine whether our earlier result depended on special features of the older formalism. We show that it does not. Although IIT3.0 offered a convenient minimal representation, the same difficulty arises in IIT4.0; the operator proliferation problem therefore reflects a structural feature of consciousness-based collapse proposals rather than an artifact of a particular formulation of IIT.

In IIT4.0 the simplest conscious system is a \emph{monad}: a single unit that feeds into itself in a self-loop, i.e., whose next state is fully determined by its current state and nothing else. To make this concrete, consider a binary unit $A$ whose next state is determined solely by its current state as happens when such a unit is sent into a NOT gate. The system is specified by a transition probability matrix (TPM), giving the probabilities of the next state $A_{t+1}$ conditional on the current state $A_t$:
$$
\begin{array}{c|cc}
 & A_{t+1}=0 & A_{t+1}=1 \\ \hline
A_t = 0 & 0 & 1 \\
A_t = 1 & 1 & 0
\end{array}
$$
Given the present state, the TPM fixes the probabilities of both past and future states of the unit. IIT4.0 evaluates this by comparing the unit’s actual causal constraints to an unconstrained 
baseline in which both past and future states are equally likely. Because the real TPM 
eliminates this uncertainty, the unit specifies its own cause and effect state with maximal 
irreducibility. This specification is called a \emph{distinction}. Thus the monad generates 
one distinction: the unit determining its own past and future state.

IIT4.0 then introduces a second level of structure. The cause and effect specifications 
produced by the mechanism are not treated as independent pieces of information; rather, the 
theory evaluates how they constrain one another. In the monad, the same unit that specifies 
a cause state also specifies a corresponding effect state, and these two specifications are 
linked by the system’s TPM. IIT quantifies this linkage as a \emph{relation}. For the NOT gate, the TPM deterministically pairs each possible past state with a corresponding future state, so the cause and effect specifications form a single irreducible cause–effect mapping relative to the unconstrained baseline. The resulting $\phi$-structure therefore consists of one distinction and one relation. The monad thus illustrates a central feature of IIT4.0: conscious structure depends on counterfactual causal organization encoded in the TPM.

Consider now the feedback dyad. Although the dyad can occupy four classical states, its causal organization, as captured by the TPM, does not vary across those states in a way that changes the structure described above. In each state the same mechanisms specify their own cause and effect constraints, and the same linkages between those specifications are present. The dyad therefore generates the same collection of distinctions and the same relations in every classical state. (Explicit calculations are provided in Appendix \ref{4.0Phi}.) Accordingly, the dyad possesses a single $\phi$-structure that is independent of whether the system is in state $(0,0)$, $(0,1)$, $(1,0)$, or $(1,1)$. What changes between these physical states are the node values themselves, not the counterfactual causal organization encoded in the TPM. Since IIT4.0 individuates conscious states by their $\phi$-structure, the dyad does not realize different experiences in its different classical states. A quantum superposition of these states would therefore not correspond to a superposition of conscious experiences.

One might think that this avoids our earlier argument. In IIT3.0 the collapse dynamics
appeared to depend directly on probability distributions represented in the Q-shape,
whereas in IIT4.0 conscious states are described instead by a $\phi$-structure consisting
of distinctions and relations. However, the relevant structure has not disappeared. The
distinctions and relations are defined through the cause and effect specifications generated
by the system, and these specifications are determined by the system’s transition probability
matrix (TPM). Consequently, the $\phi$-structure still depends on how the system would behave
under counterfactual alternatives encoded in the TPM. A collapse model that makes collapse rates depend on conscious structure must therefore be sensitive to this counterfactual causal organization. The dependence is no longer expressed
through explicit probability distributions appearing in the description of the conscious
state, but it remains implicitly present through the TPM from which the $\phi$-structure is
derived.

This has a direct consequence for the collapse dynamics. Earlier we showed that a single
collapse operator cannot in general make the collapse rate depend only on conscious
experience. Any such theory must therefore employ multiple operators if it is to track the
features that individuate experiences. In IIT4.0, experiences are specified by a collection of distinctions and relations.
Changing one relation while leaving the others unchanged constitutes a change in
experience according to the theory. A collapse model whose rates depend on conscious
structure must therefore be able to respond to these independent features. It follows that
separate collapse operators are required for the independent relations that make up the
$\phi$-structure. The number of operators required by such a theory therefore grows with
the number of relations present in the system.

The scaling becomes dramatic very quickly. Even very small IIT4.0 systems possess many
relations. The three-node system illustrated in Figure~8A of \cite{albantakis2022integrated}
contains 60 relations, while the five-node system of Figure~7A contains more than 13,000.
This proliferation does not arise from unusually complicated wiring. Rather, it follows
from how IIT4.0 defines structure. Any subset of nodes can function as a mechanism and generate a distinction, and relations arise when multiple mechanisms jointly constrain the same past or future states. As the number of nodes increases, the number
of possible mechanisms grows exponentially, and the number of ways their constraints can
overlap grows combinatorially. A collapse model that tracks conscious structure would therefore require at least a corresponding number of independent collapse operators. The resulting dynamics becomes
extremely complex even for very small systems, challenging the suggestion that
IIT-based collapse models would be especially simple or experimentally tractable.

To see why  operator proliferation might make the experimental testing of consciousness-based collapse models hard, consider the following aspects of such tests. An important challenge will be to delineate proper consciousness-induced collapse from other mechanisms such as environment-induced decoherence or collapse due to gravitational interaction. For this distinction, it is important to have concrete predictions about how collapse rates depend on the structure of the system and its associated conscious experience, rather than simply postulating and perhaps observing a certain decoherence time scale. It is reasonable to expect that consciousness-induced collapse is only practically observable in systems that are sufficiently complex. Computing the Q-shapes and their distances for such systems is expected to be computationally very difficult, due to the involved combinatorial complexities. On the one hand, this makes the actual prediction of concrete collapse times difficult; on the other hand, it may suggest that nature rather implements an incomplete correspondence between collapse rates and differences of conscious experience, similar to the one we have seen in the dyad for a single collapse operator. (Indeed, the impossibility to use natural systems for the solution of computationally very complex problems has been suggested as a physical principle~\cite{Aaronson2005}, similar to other impossibility principles such as no-cloning or the non-existence of a perpetuum mobile in thermodynamics). But if this is true, then experiments cannot simply test whether experientially more distinct states of consciousness lead to faster decoherence, and it is unclear how proper consciousness-induced collapse should be structurally distinguished from, say, more ordinary environment-induced decoherence.

The robustness of our conclusion is worth emphasizing. We have shown that the operator
proliferation problem arises in (a version of) IIT3.0 and persists under the substantially revised
formalism of IIT4.0. The difficulty therefore does not depend on the particular way in
which conscious structure is represented—whether by Q-shapes or by $\phi$-structures —
but on the more basic requirement that collapse rates track the structure of experience
rather than solely on properties of the physical state, such as mass-density distributions.

Indeed, the underlying tension may extend beyond IIT altogether. Any theory of
consciousness that individuates experiences in terms of rich internal organization—for
example, by appeal to patterns of connectivity, causal dependence, or counterfactual
structure in addition to momentary activation states—will generate multiple independent
features that distinguish one experience from another. This idea is not specific to IIT and is taken seriously even by some critics of IIT, for example, the claim in \cite{fleming2024quality} that ``merely possible experiences should affect phenomenal
character''. A collapse model that attempts to
make its dynamics sensitive to such features will therefore require correspondingly many
independent dynamical parameters. The proliferation we have identified thus reflects a
general structural pressure on Wigner-style consciousness-based collapse proposals, rather
than a peculiarity of IIT.

\section{Conclusions}

In this paper we examined whether Wigner-style consciousness-causes-collapse theories yield simple, testable predictions, when combined with mathematical theories of consciousness such as integrated information theory (IIT). Using the feedback dyad as a minimal and fully explicit model system, we analyzed both an IIT-based characterization of conscious states and the dynamical constraints faced by simple consciousness-based collapse models.

Our central technical result is a structural constraint on collapse models of the Lindblad form considered here: if collapse is governed by too few collapse operators (most notably, a single operator) then collapse rates cannot, in general, be made to depend only on pairwise qualitative differences between conscious states. We also showed that this limitation can, in principle, be avoided by introducing sufficiently many commuting collapse operators, but only at the cost of a rapid proliferation of operators even for very simple systems.

This trade-off bears directly on the claim that IIT-based consciousness-collapse models offer especially simple or experimentally tractable alternatives to standard collapse theories. While such models are not ruled out by our results, the complexity of the collapse dynamics itself becomes a central obstacle to assessing their empirical viability. Classical IIT already suffices to generate this difficulty in concrete systems, and quantum extensions of IIT substantially increase the complexity of the resulting models without alleviating the structural constraints identified here.

We further showed that the operator-proliferation problem is not an artifact of the particular IIT3.0 formalism used in the original collapse proposal. Repeating the analysis using IIT4.0, where conscious states are represented by $\phi$-structures rather than Q-shapes, leads to the same difficulty in a different form. In IIT4.0 the relevant structure includes relations linking different experiential distinctions, and the number of such relations grows rapidly with system size. A collapse dynamics whose rates track conscious structure must therefore incorporate a corresponding number of dynamical terms. The resulting proliferation follows from how the theory individuates experiences, not from the details of any specific representation.

This suggests a more general moral. The source of the difficulty is that collapse rates are required to depend on differences between experiences, while experiences—on IIT and on many other approaches to consciousness—are individuated partly by counterfactual causal structure, not merely by its instantaneous physical state. Any theory of consciousness that takes the connectivity or causal organization of a system (for example, neuronal connectivity in addition to activation state) to be relevant to experience will introduce multiple independent features distinguishing one experience from another. A collapse dynamics sensitive to those differences must therefore contain correspondingly many independent dynamical parameters.

Taken together, our results identify constraints on consciousness-based collapse models that arise from the general structure of quantum theory. They do not show that consciousness-based collapse is impossible. Rather, they show that once collapse rates are required to track structured conscious experience, the resulting dynamics becomes highly constrained and generically complex. This helps to delineate the space of viable consciousness-based collapse proposals and the challenges they face as candidates for empirically testable modifications of quantum mechanics.

\vspace{6pt}

\acknowledgments{We are grateful to Larissa Albantakis for extensive feedback on earlier drafts, to Thomas D.\ Galley for helpful discussions, and to an anonymous referee. This research was supported by grant number FQXi-RFP-CPW-2015 from the Foundational Questions Institute and Fetzer Franklin Fund, a donor advised fund of Silicon Valley Community Foundation. This project/publication was also made possible through the support of Grant 63209 from the John Templeton Foundation. The opinions expressed in this publication are those of the authors and do not necessarily reflect the views of the John Templeton Foundation. Moreover, this research was supported in part by Perimeter Institute for Theoretical Physics. Research at Perimeter Institute is supported by the Government of Canada through the Department of Innovation, Science, and Economic Development, and by the Province of Ontario through the Ministry of Colleges and Universities.
}

\appendix

\section[\appendixname~\thesection]{Solution of the Optimization Problem of Section~\ref{collapse}}\label{optimal}

First note that replacing $\lambda_{ij}$ by $\lambda'_{ij}:=\lambda_{ij}-m$, where $m:=\min_{k,l}\lambda_{k,l}$, will also satisfy the constraints, but lead to $\lambda'_{00}+\lambda'_{01}+\lambda'_{10}+\lambda'_{11}< \lambda_{00}+\lambda_{01}+\lambda_{10}+\lambda_{11}$ unless $m=0$. Thus, to be a minimizer, at least one of the $\lambda_{ij}$ must be zero.

Consider, for example, the case that \mbox{$\lambda_{01}=0$. Then $|\lambda_{00}-\lambda_{01}|\geq \mathcal{D}(Q(0,0),Q(0,1))=2$,} and so $\lambda_{00}\geq 2$. Similarly, $|\lambda_{10}-\lambda_{01}|\geq\mathcal{D}(Q(1,0),Q(0,1))=4$, and so $\lambda_{10}\geq 4$. Finally, $|\lambda_{11}-\lambda_{00}|\geq \mathcal{D}(Q(1,1),Q(0,0))=4$, and so either $\lambda_{11}\geq 4+\lambda_{00}$ or $\lambda_{00}\geq 4+\lambda_{11}$. For the ``either'' case, the smallest possible choices of the eigenvalues are $(\lambda_{00},\lambda_{01},\lambda_{10},\lambda_{11})=(2,0,4,6)$, and it is easy to check that this also satisfies all other desired inequalities. For the ``or'' case, consider the additional inequality $|\lambda_{11}-\lambda_{01}|\geq\mathcal{D}(Q(1,1),Q(0,1))=2$, i.e.,\ $\lambda_{11}\geq 2$, which gives us another solution $(6,0,4,2)$.

All other minimizers can be determined similarly, but considering all the other three cases of zero $\lambda_{ij}$.

\section[\appendixname~\thesection]{IIT4.0 Calculation of Dyad $\Phi$ and $\phi$-Structure}\label{4.0Phi}

This section uses IIT4.0 \cite{albantakis2022integrated} to calculate the dyad's amount of consciousness ($\Phi$) and to model its state of consciousness ($\phi$-structure). The general procedure for calculating $\Phi$ takes several steps. Fortunately, the simplicity of our dyad allows us to skip some of these and to emphasize the most important ones. We refer the reader to~\cite{albantakis2022integrated} for a more general presentation and for detailed definitions of the terminology that we are using.

The dyad consists of two parts, A and B. We begin by calculating the integrated cause and effect information specified by each unit in its current state. In IIT4.0 a unit in a given state specifies a distinction, and its integrated information is the irreducibility of that distinction. Integrated information concerns how much information is lost by partitioning the system, which means replacing a causal relationship with noise where the noise is represented as an equiprobable distribution over all possible states. 

To illustrate, let us calculate how much integrated effect information A has, given its present state, about the next state of each of the system's parts, A and B. The maximum of these defines A's integrated effect information. 

Given that our dyad is a SWAP gate, it is trivially true that A's present state has zero integrated effect information about A's next state since A's next state is entirely determined by B's present state. Put another way, A's possible next states are all equally probable given its current state. So introducing a partition that induces noise between A at $t_0$ and A at $t_{+1}$ makes no difference. This makes sense given that there is no causal connection between them in the first place: A affects B but not itself in the next time step. A's present state is not causally connected to A's future state and so there is no integrated effect information.

However, A's present state \textit{does} fully determine B's future state and so if, for example, our system's present state is (1,0), Equation~(41) in \cite{albantakis2022integrated} tells us that the integrated effect information of A's state at time $t_0$ given that it is in state 1 at that instant is
\begin{equation}
\phi_e(a_{t_0}=1) = p(b_{t_{+1}}=1 | a_{t_0}=1) \log_2 \left[ \frac{p(b_{t_{+1}}=1 | a_{t_0}=1)}{p^\theta(b_{t_{+1}}=1 | a_{t_0}=\textrm{noise})} \right].
\label{phi:e}
\end{equation}
Here, $p(b_{t_{+1}}=1 | a_{t_0}=1)$ is the probability that B will be in state $b=1$ at time $t_{+1}$ given that A is currently in state 1. It is trivially true that this equals 1. Likewise \mbox{$p^\theta(b_{t_{+1}}=1 | a_{t_0}=\textrm{noise})$} represents the probability that B will be in state 1 at time $t_{+1}$ given the partition $\theta$ which sets the value of channel A to an equiprobable distribution of the two possible states. In other words, the partition replaces the effect that A had on B with noise, which means that B's future state is randomly determined. Since there are only two possible states, that means that $p^\theta(b_{t_{+1}}=1 | a_{t_0}=\textrm{noise})=0.5$. As such, we have
\begin{equation}
\phi_e(a_{t_0}=1) = 1 \cdot \log_2 \left[ \frac{1}{0.5} \right] = 1.
\label{phi:e:calc}
\end{equation}
The same basic equation tells us that the integrated effect information of B's state at time $t_0$, $\phi_e(b_{t_0}=0)$, also equals 1.

The integrated cause information for A is calculated in a slightly different manner and illustrates a time asymmetry in the equations of IIT. As in the effect case, the past state of A contains no information about the present state of A, and likewise for B. We only consider the information B's past state has on A's current state and the information A's past state has on B's current state. Specifically, given a current state of (1,0), Equation~(42) in \cite{albantakis2022integrated} gives
\begin{equation}
\phi_c(a_{t_0}=1) = p(b_{t_{-1}}=1|a_{t_0}=1)\log_2\left[\frac{p(a_{t_{0}}=1|b_{t_{-1}}=1)}{p^{\theta}(a_{t_{0}}=1|b_{t_{-1}}=\textrm{noise})}\right]
\label{phi:c}
\end{equation}
where $p(b_{t_{-1}}=1|a_{t_0}=1)$ is calculated according to Bayes' rule as follows:
\begin{equation}
p(b_{t_{-1}}=1|a_{t_0}=1) = \frac{p(a_{t_{0}}=1|b_{t_{-1}}=1)\cdot p(b_{t_{-1}}=1)}{p(a_{t_{0}}=1)}
\label{bayes}
\end{equation}
where $p(a_{t_{0}}=1)$ and $p(b_{t_{-1}}=1)$ are unconstrained probabilities\mbox{ (see Equations (6)--(8) in \cite{albantakis2022integrated}),} which IIT4.0 takes to be uniform over past states (maximum entropy), and therefore equal to 0.5. Here we also have that $p(a_{t_{0}}=1|b_{t_{-1}}=1)$ is the probability that A's current state is 1 if B's past state is 1 and $p(b_{t_{-1}}=1|a_{t_0}=1)$ is the probability that B's past state was 1 given that A's state is currently 1. As before, $p^{\theta}(a_{t_{0}}=1|b_{t_{-1}}=\textrm{noise})$ noises the system and is equal to 0.5. Since $p(a_{t_{0}}=1|b_{t_{-1}}=1)=1$, Bayes' rule given by Equation~(\ref{bayes}) tells us that $p(b_{t_{-1}}=1|a_{t_0}=1)=1$. As before, then, we find that $\phi_c(a_{t_0}=1)=1$. Likewise, the same process tells us that $\phi_c(b_{t_0}=0)$ also equals 1.

Equation (45) in \cite{albantakis2022integrated} then tells us that the integrated information of a part is the minimum of its integrated effect and integrated cause information, i.e.,
\begin{align}
\phi(a_{t_0}=1) & = \min\left[\phi_c(a_{t_0}=1),\phi_e(a_{t_0}=1)\right] \\ \phi(b_{t_0}=0) & = \min\left[\phi_c(b_{t_0}=0),\phi_e(b_{t_0}=0)\right]
\label{min:ii}
\end{align}
respectively, which are both trivially 1. Each unit therefore specifies one irreducible distinction of unit value.

The calculations above determine the integrated information of the dyad’s two
\emph{distinctions}: in the state $(1,0)$, unit $A$ specifies that $B$
was $1$ at $t_{-1}$ and will be $1$ at $t_{+1}$, while unit $B$
specifies that $A$ was $0$ at $t_{-1}$ and will be $0$ at $t_{+1}$.
In IIT4.0 the $\phi$-structure also includes \emph{relations} between
distinctions (see \cite{albantakis2022integrated} for the general
definition). In the present case, each distinction links a statement
about the past to a corresponding statement about the future, and this
link depends on the dyad’s transition rule. If the connection between
the units were replaced by noise (as in the relevant partition), that
link would disappear. The strength of this linkage is what IIT4.0
quantifies as a relation.

In the SWAP dyad this linkage is maximal. Consider the distinction generated by $A$.
Given $a_{t_0}=1$, the transition rule implies both
$p(b_{t_{-1}}=1\mid a_{t_0}=1)=1$ and
$p(b_{t_{+1}}=1\mid a_{t_0}=1)=1$:
$A$ fixes both the past and the future state of $B$.
Under the relevant partition, however, the connection from $A$ to $B$ is replaced
by uniform noise, so the corresponding probabilities become $1/2$.
The linkage between the past and future specifications is therefore irreducible
relative to the partitioned baseline by
\begin{equation}
\phi_r(A)=\log_2\!\left(\frac{1}{1/2}\right)=1.
\label{rel:A}
\end{equation}
By symmetry, the same reasoning yields $\phi_r(B)=1$.

The dyad’s $\phi$-structure therefore consists of two distinctions and two relations,
each with unit value. Accordingly, the total amount of consciousness is
\begin{equation}\label{PhiDyadIIT4}
\Phi(t_0)=\underbrace{\phi_d(A)+\phi_d(B)}_{\text{distinctions}}+
\underbrace{\phi_r(A)+\phi_r(B)}_{\text{relations}}
=1+1+1+1=4.
\end{equation}
No matter which of its four possible classical states $((0,0),(0,1),(1,0),(1,1))$
the dyad occupies, the same pattern of constraints holds, and hence the same
$\phi$-structure and the same value of $\Phi$ are obtained.

\end{document}